\documentclass{iaus}
\usepackage{graphicx}
\usepackage{wasysym}

\title[Where is the snow line?] %% give here short title %%
{The Location of the Snow Line in Protostellar Disks}

\author[Morris Podolak]   %% give here short author list %%
{Morris Podolak}

\affiliation{Dept. of Geophysics and Planetary Sciences, Tel Aviv University, \\ Tel Aviv,
Israel \\ email: {\tt morris@post.tau.ac.il} \\[\affilskip]}

\pubyear{2009}
\volume{263}  %% insert here IAU Symposium No.
\pagerange{119--126}
\jname{Icy Bodies of the Solar System}
\editors{D. Lazzaro, D. Prialnik, R. Schulz \& J.A. Fernandez, eds.}
\begin{document}

\maketitle

\begin{abstract}
The snow line in a gas disk is defined as the distance from the star beyond which the water ice is stable against evaporation.  Since oxygen is the most abundant element after hydrogen and helium, the presence of ice grains can have important consequences for disk evolution.  However, determining the position of the snow line is not simple.  I discuss some of the important processes that affect the position of the snow line. 

\end{abstract}

\firstsection % if your document starts with a section,
\section{Introduction}

Oxygen is the most abundant element after hydrogen and helium, and water is one of the most abundant oxygen-bearing molecules.  When water is a vapor, the mass fraction of solids in a solar composition disk is $\sim 4.2\times 10^{-3}$.  After water condenses, the solid mass fraction jumps to $\sim 1.3\times 10^{-2}$ \cite[(Anders \& Grevasse 1989)]{AndersGrevasse89}.  The properties of a protostellar disk will therefore change significantly when we cross the boundary between water vapor and ice grains.  These changes can have important consequences for other processes taking place in the disk.  

\begin{figure}
 \vspace*{2.0 cm}
\begin{center}
 \includegraphics[width=4in]{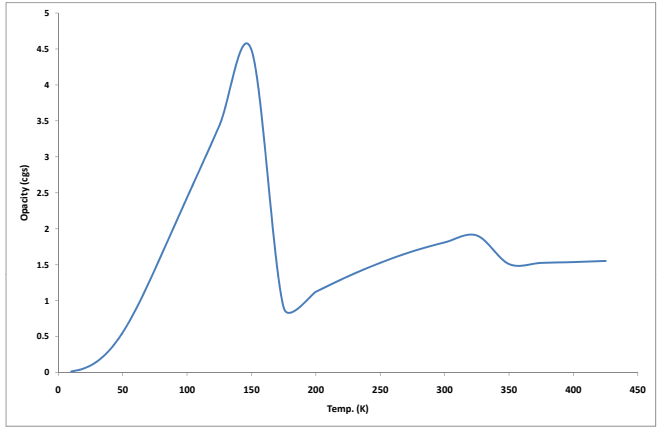} 
 \vspace*{-2.0 cm}
 \caption{Opacity of dust grains thought to have been present in the early solar nebula as a function of temperature.  The opacities are taken from \cite[Pollack et al. (1985)]{Polletal85}.}
   \label{pollakopac}
\end{center}
\end{figure}

One consequence is the change in opacity that accompanies the transition from a region where water is a vapor to a region where it is a solid.  This is illustrated in Fig.\,\ref{pollakopac} which is based on the Rosseland mean opacities of \cite[Pollack et al. (1985)]{Polletal85}.  
At temperatures above around 170 K, the opacity is due to mainly rocky grains, but just below this temperature the opacity jumps by nearly a factor of 5 because of the sudden condensation of water vapor to ice grains.  At still lower temperatures the opacity again drops because, in taking the Rosseland mean for low temperatures, the longer wavelengths are given more weight.  At these longer wavelengths the ice grains are inefficient scatterers and absorbers, so the opacity decreases.  Such changes in opacity can have an important effect on the evolution of the gas disk.

Another example of the possible significance of the snow line is the mechanism suggested by Stevenson and Lunine \cite[(1988)]{StevLun88} for forming Jupiter.  Sunward of the snow line the disk temperature is high enough so that all the water is in the vapor phase.  Once you cross the snow line and water begins to condense, the mass fraction of water in the vapor phase decreases.  This causes a composition gradient across the snow line which drives diffusion of water vapor into the colder region and causes this region to become enhanced in solids.  Stevenson and Lunine \cite[(1988)]{StevLun88} argued that this enhancement could lead to the formation of Jupiter.  The problem with this scenario is that this mechanism forms only one giant planet.  Some other mechanism would be required for Saturn.  But if it is correct, then the snow line had to have been somewhere in the vicinity of Jupiter, i.e. at around 5 AU.

Unfortunately, Jupiter appears to be sending us contradictory messages.  An analysis of Jupiter's composition by Lodders \cite[(2004)]{Lodders04} finds that Jupiter is anomalously low in water, and concludes that the snow line was much further away from the sun; perhaps in the neighborhood of 10 AU.

A third reason for trying to determine the position of the snow line is that it tells us where we can expect to see icy bodies.  This has particular significance in view of the recent discovery of main-belt comets \cite[(Hsieh and Jewitt 2008)]{HsieJewit08}.  These bodies have orbits that are dynamically similar to those of the main belt asteroids \cite[(Jewitt et al. 2009)]{Jewitetal09} yet display cometary activity.  If main-belt comets were indeed formed in the asteroid belt, the snow line must have been closer to the sun than around 3 AU.  

\section{Theoretical Calculation of the Snow Line}
What can theory tell us?  One early attempt to compute the position of the snow line was by Hayashi \cite[(1981)]{Hayashi}.  He argued that the stellar flux should decrease as $r^{-2}$ where $r$ is the distance from the star, so that the temperature at some point in the disk, $T(r)$ should be given by
\begin{equation}\label{eq:hayash}
\sigma T(r)^4=\frac{1}{4}\sigma T_*^4\left(\frac{R_*}{r}\right)^2
\end{equation}
Here $R_*$ is the stellar radius, $T_*$ is the stellar surface temperature, and $\sigma$ is the Stefan-Boltzmann constant.  For a solar composition gas at a pressure typical for protostellar disks the temperature at which water starts to condense is roughly 170 K.  Setting $T(r)=170$\,K and using the solar radius and temperature in eq.\,(\ref{eq:hayash}) gives $r=2.7$\,AU.  This is very close to the value needed to explain the main-belt comets.

The problem is that Hayashi's model is too simple.  More modern views of the protostellar disk see it as physically thin, and optically thick.  In the case of a {\it passive disk} with no internal heat sources, the temperature is determined by integrating the flux falling on a point in the disk from different parts of the stellar surface.  The resulting temperature distribution is [see, e.g. \cite[Armitage (2009)]{Armitage09}]
\begin{equation}\label{eq:fltdsk}
\sigma T^4_{rad}(r)=\frac{1}{\pi}\sigma T_*^4\left[\sin^{-1}\left(\frac{R_*}{r}\right)-\left(\frac{R_*}{r}\right)\sqrt{1- \left(\frac{R_*}{r}\right)^2}\right]\approx \frac{2}{3\pi}\sigma T_*^4\left(\frac{R_*}{r}\right)^3
\end{equation}
for $r\gg R_*$.  Since this temperature is due to radiative heating I have denoted it by $T_{rad}(r)$

This expression too must be modified to allow for the fact that the disk is flared, and the outer regions present a more direct face to the star and therefore absorb more light.  As a result the flaring is even greater, and the coefficient $2/(3\pi)$ in eq.\,\ref{eq:fltdsk} must be replaced by some function of $r$.  Finally, if the central star is accreting material from the disk, there will be a flux of material moving through the disk which will provide a source of viscous heating.  This provides an additional flux that can be written as \cite[(Armitage 2009)]{Armitage09} 
\begin{equation}
\sigma T_{visc}^4(r)=\frac{3GM_*\dot{M}}{8\pi r^3}\left(1-\sqrt{\frac{R_*}{r}}\right)
\end{equation}
where $M_*$ is the mass of the central star, $\dot{M}$ is the mass accretion rate, and $G$ is Newton's constant.  The actual temperature profile in the disk will be given by the sum of the radiative and viscous fluxes
\begin{equation}\label{tprof}
\sigma T^4(r)=\sigma T_{rad}^4(r)+\sigma T_{visc}^4(r)
\end{equation}

\cite[Sasselov \& Lecar (2000)]{SasLec00} solved eq.\,(\ref{tprof}) numerically, and found that the snow line fell between 0.7 and 1.5\,AU, depending on the strength of the accretional heating.  While this allows {\it in situ} formation of the main-belt comets, it also implies that {\it all} the main-belt asteroids should be icy bodies.  This would appear to put the snow line too close to the sun.  

\section{Grain Thermal Balance}
Although the model of Sasselov \& Lecar (2000) does a much better job of calculating the temperature of the disk, is still assumes that the temperature at the snow line is 170\,K.  The actual condensation temperature of water vapor will depend on the partial pressure of water vapor in the disk, and this will vary from place to place.  In addition, the temperature of an ice grain will not necessarily be the same as the temperature of the ambient gas.  Rather, it will depend on the details of energy balance with the surrounding medium.  This energy balance is encompassed in the three heating and two cooling terms given below [see \cite[Mekler \& Podolak (1994)]{MekPod94} and \cite[Podolak \& Mekler (1997)]{PodMek97} for further details].  In all cases the heating and cooling are per unit area of the grain.\bigskip\\

\underline{Radiative Heating} - The heating of the grain by radiation from the star (for the optically thin case) or the surrounding medium (for the optically thick case).  Here $a$ is the grain radius, $\lambda$ is the wavelength of the radiation, and $S(\lambda)$ is the wavelength dependent radiation field. $Q_{abs}$ is the efficiency of absorption per unit area and is a function of $a$, $\lambda$, and the grain material.  It can be computed from Mie theory \cite[(van de Hulst 1957)]{vdHulst57}.
\begin{equation}
E^h_{rad}=\frac{1}{4r^2}\int_0^{\infty}Q_{abs}(a,\lambda)S(\lambda)d\lambda
\end{equation}

\underline{Gas Heating} - The heating of the grain by contact with the gas.  Here $n_{H_2}$ is the number density of hydrogen molecules in the gas, $m_{H_2}$ is the mass of a hydrogen molecule, $T_{gas}$ and $T_{grain}$ are the gas and grain temperatures, respectively, $j$ is the number of thermodynamic degrees of freedom of the hydrogen molecule, and $k$ is Boltzmann's constant.
\begin{equation}
E^h_{gas}=\frac{n_{H_2}}{4}\sqrt{\frac{8kT_{gas}}{\pi m_{H_2}}}\frac{jk(T_{gas}-T_{grain})}{2}
\end{equation}
Note that if the grain is warmer than the gas, this term is negative.

\underline{Condensation Heating} - The heating of the grain due to water condensing on it.  Here $n_{H_2O}$ is the number density of water molecules in the gas, $m_{H_2O}$ is the mass of a water molecule, and $q$ is the latent heat released when a water molecule condenses onto the ice grain.
\begin{equation}
E^h_{cond}=\frac{n_{H_2O}}{4}q\sqrt{\frac{8kT_{gas}}{\pi m_{H_2O}}}
\end{equation}

\underline{Radiative Cooling} - The cooling of the grain by radiation to space.  Here $Q_{emis}=Q_{abs}$ is the efficiency factor for emission of radiation, and $F$ is the Planck function at the temperature of the grain.
\begin{equation}
E^c_{rad}=\int_0^{\infty}Q_{emis}(a,\lambda)F(\lambda,T_{grain})d\lambda
\end{equation}

\underline{Evaporative Cooling} - The cooling of the grain by water sublimation.  Here $P_{vap}$ is the vapor pressure of water.
\begin{equation}
E^c_{evap}=q\frac{P_{vap}(T_{grain})}{\sqrt{\pi m_{H_2O}kT_{grain}}}
\end{equation}

\begin{figure}
 \vspace*{2.0 cm}
\begin{center}
 \includegraphics[width=10cm]{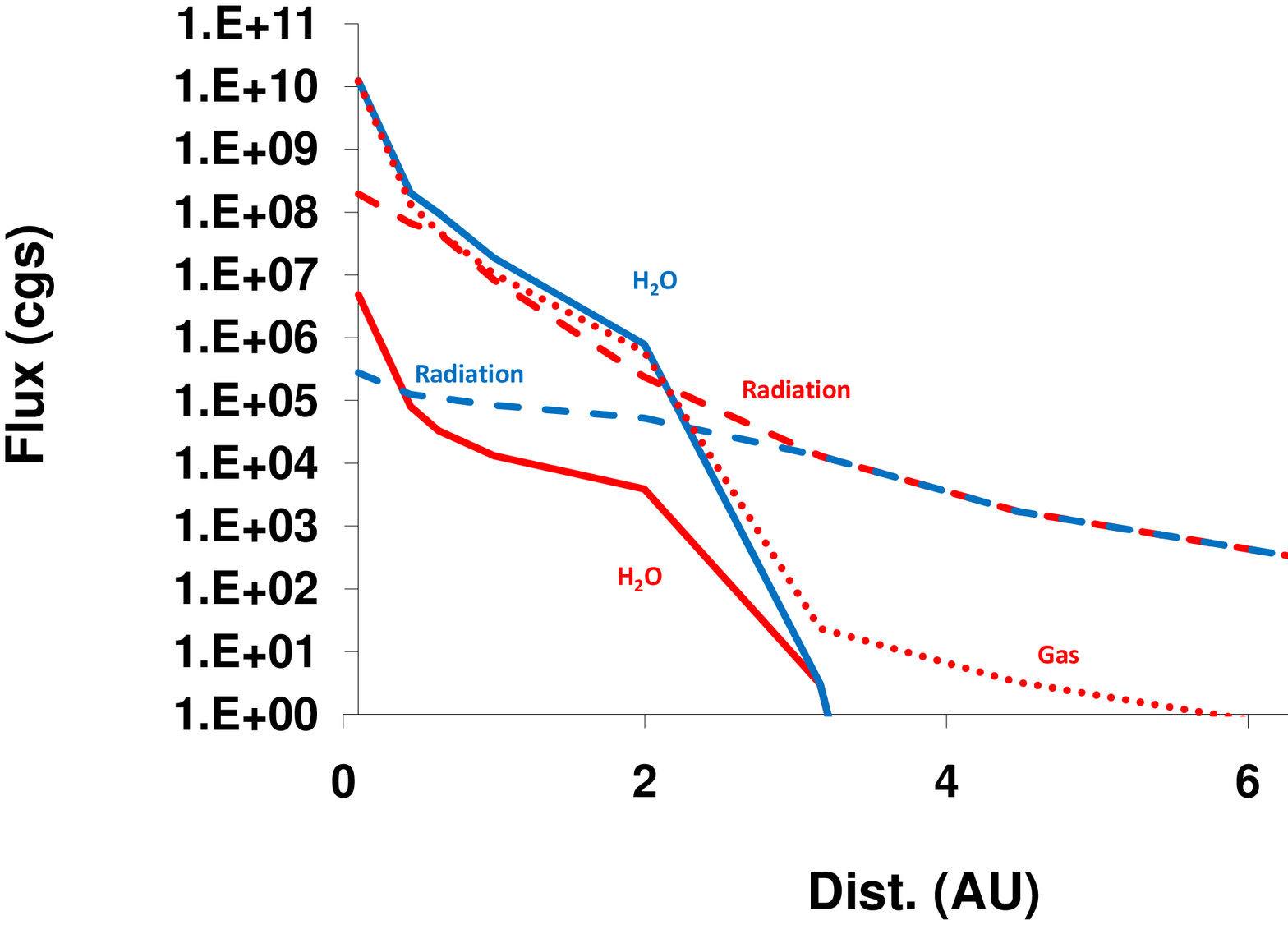}
\includegraphics[width=10cm]{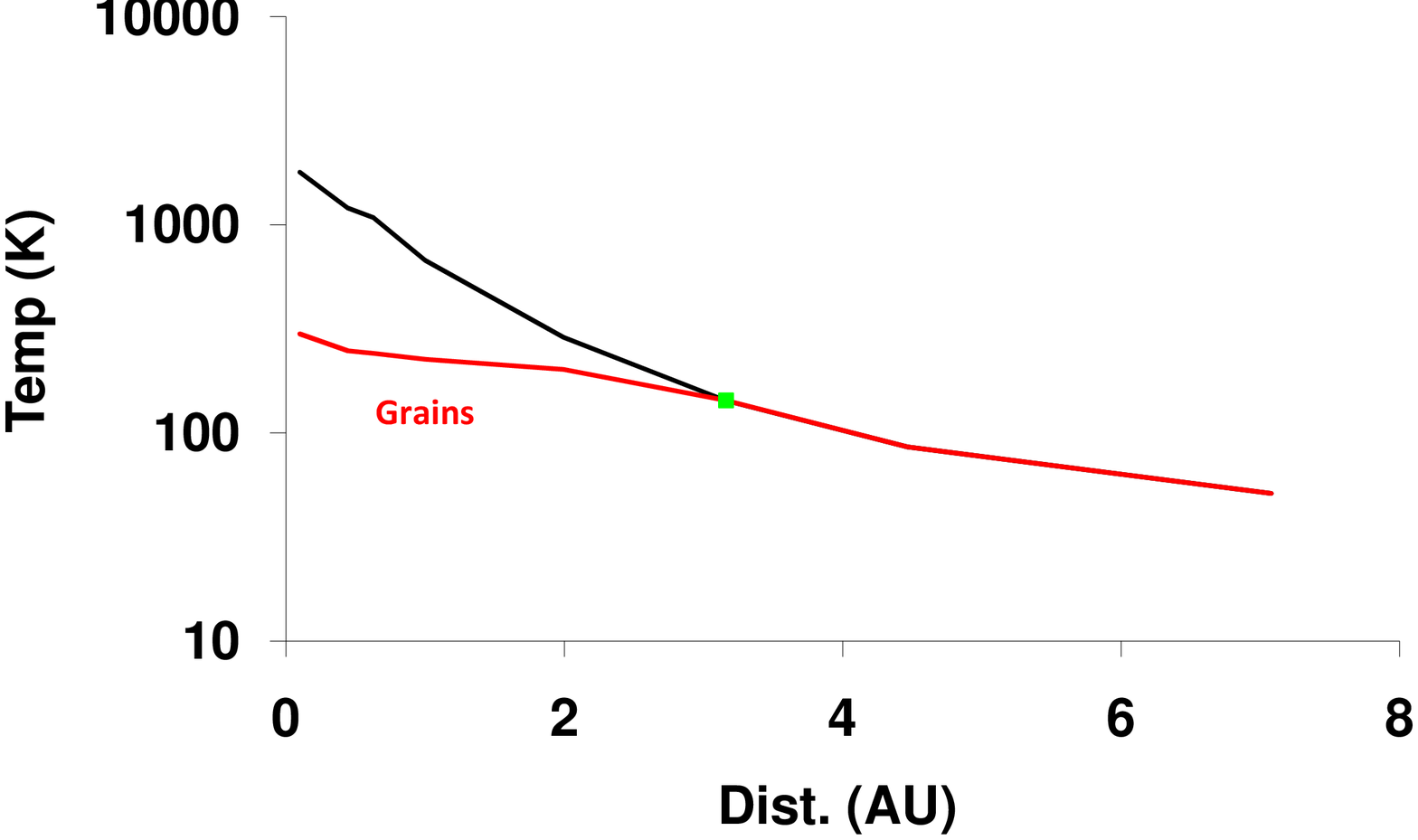} 
 \vspace*{-3.0 cm}
 \caption{The upper panel shows the heating (red) and cooling (blue) terms for a 10 $\mu$m grain in an optically thick gas disk as a function of distance from the star. Water condensation and evaporation (solid curves), radiative heating and cooling (dashed curves) and energy transfer by contact with the gas (dotted curve) are shown.  The lower panel shows the temperatures of the gas (black) and grains (red).  The green square indicates the position of the snow line.}
   \label{midpln}
\end{center}
\end{figure}

In steady state the heating and cooling terms must be equal, and we can use the above equations to find $T_{grain}$.  The fluxes due to the different terms are shown in Fig. \ref{midpln} for the case of a 10\,$\mu$m pure ice grain in an optically thick disk.  

The disk model used in these calculations is taken from the work of \cite[Bell et al. (1997)]{Belletal97} and corresponds to the model parameters $\alpha=10^{-4}$ and $\dot{M}=10^{-8}$.  We can see that near the star the major heating term is contact with the gas, and this is offset by evaporative cooling.  The grains are kept much cooler than the gas, and they evaporate in a very short time. Further from the star the radiative heating becomes important, but evaporation is still the major cooling term.  Since evaporative cooling is strongly dependent on grain temperature, it drops quickly with distance from the star, and by around 2.5 AU radiative cooling becomes the dominant cooling mechanism.  At around 3.5 AU the evaporative flux equals the condensation flux, and the snow line is reached.

Because the temperatures near the snow line are low, the Planck peak of the radiation is in the tens of microns range, and is large compared to the radius of the grain.  As a result grains of 10\,$\mu$m and smaller are small compared to the wavelength and are inefficient both at emission and absorption.  In addition, since most of the radiation is in the far IR, where water has strong absorption features, pure ice grains will behave pretty much like grains composed of a mixture of water and some darker material (``dirt").  As a result, the position of the snow line is quite insensitive to the composition or size of the grains.  However, because the flux of condensing vapor depends on the pressure of the background gas, the temperature of the snow line will vary as a function of temperature as well.  Depending on the density, the snow line temperature can be between 150 and 180\,K.  Based on these arguments, \cite[Lecar et al. 2006]{Lecaretal06} recomputed a disk model and found that the snow line lies in the neighborhood of 2\,AU, although it can be moved out as far as 3\,AU if the opacity is increased by a factor of 5.

\section{The Photospheric Snow Line}
There is an additional way to determine the position of the snow line, and that is by observations of disks around other stars.  Such work is just beginning, and recently ice was detected in a disk around HD 142527 \cite[(Honda et al. 2009)]{Hondaetal09}.  Here energy balance considerations are even more important.  Observations see the grains near the surface of the disk, but because these grains are near the surface, they are exposed to the direct radiation from the star.

\begin{figure}
 \vspace*{2.0 cm}
\begin{center}
 \includegraphics[width=10cm]{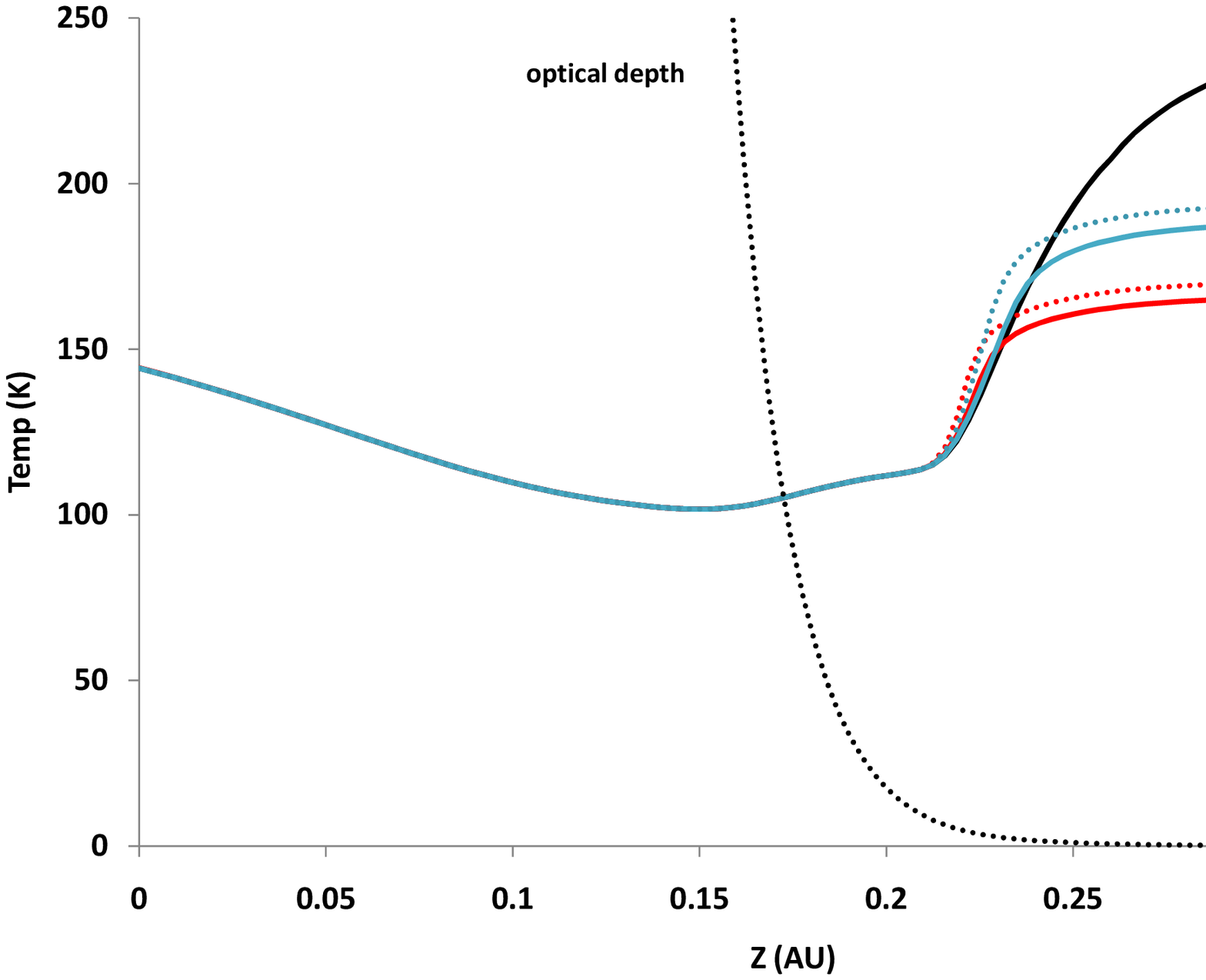}
\includegraphics[width=10cm]{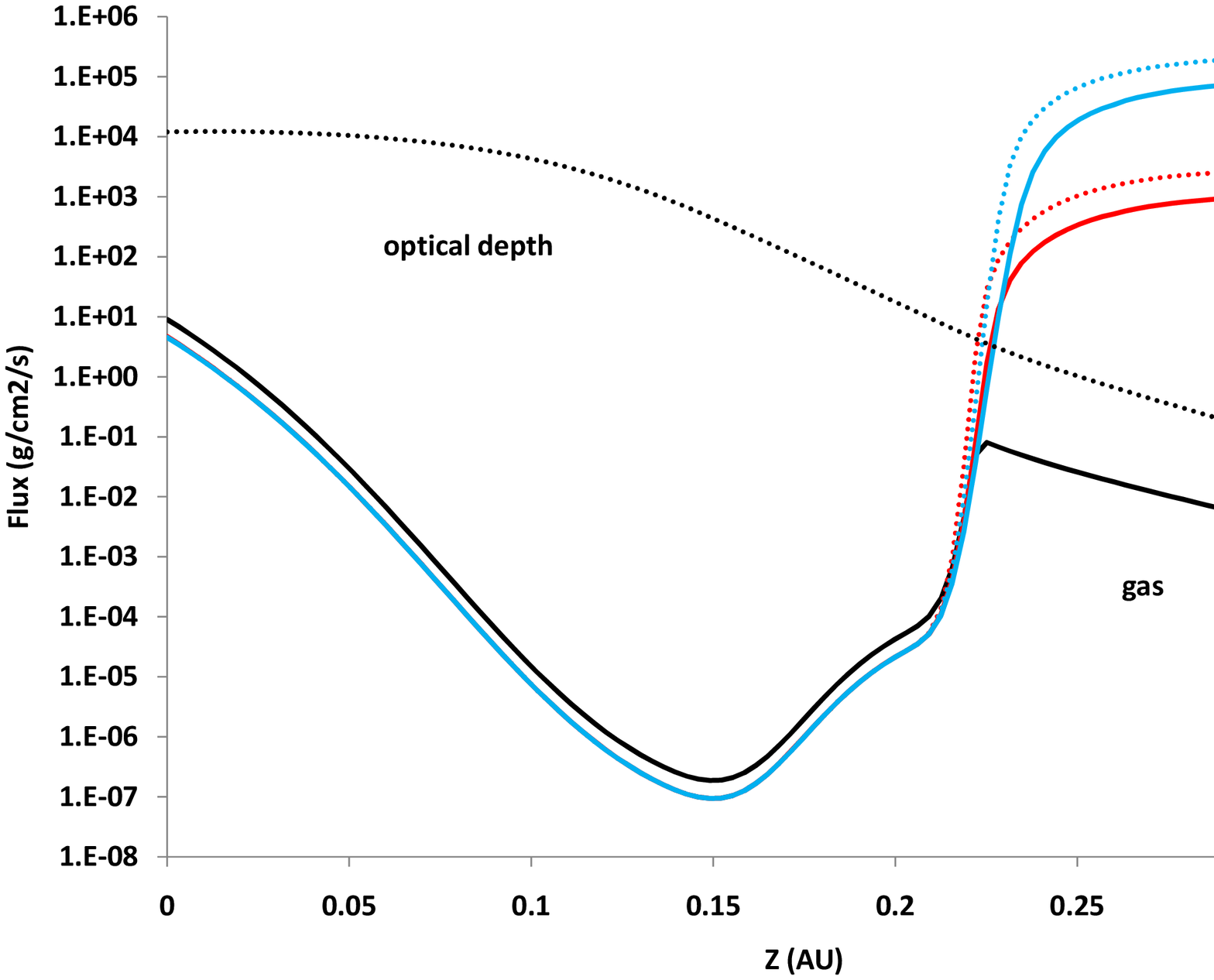} 
 \vspace*{-3.0 cm}
 \caption{The upper panel shows the temperature of 10\,$\mu$m (blue) and 0.1\,$\mu$m (red) grains as a function of height above midplane at a radial distance of 1.57 AU from the star. The solid curves are for pure ice grains and the dotted curves are for grains with 10\% by mass of generic dark material.  The solid black curve is the gas temperature.  The dotted black curve is the optical depth to the star in the visible.  The disk model is from \cite[Jang-Condell (2008)]{Jangc08}.  The lower panel shows the flux of water evaporating off the grains.  The solid black curve is the condensation flux onto the grains.}
   \label{tphot}
\end{center}
\end{figure}

Unlike the optically thick case discussed above, here the size and composition of the grains are much more important. In particular, 0.1\,$\mu$m grains will have a circumference roughly equal to the wavelength of the impinging stellar radiation, and will therefore interact with it efficiently.  However, at the relevant temperatures the wavelength of the emitted radiation will be orders of magnitude larger and will not be emitted efficiently.  In addition, since water is transparent in the visible, the absorption cross section of the grains to visible radiation will differ significantly between pure ice grains and dirty ice grains.

These effects can be seen clearly in Fig. \ref{tphot}.  For both large and small grains composed of either pure or dirty ice, the grain temperature follows the gas temperature closely so long as the optical depth to the star is high.  Once the optical depth drops to order one, the grain temperatures differ from the gas temperatures, and the grain size and composition make a difference.  In fact, the grains are sometimes warmer and sometimes cooler than the surrounding gas.  There are several things worth noting here:  First, once the visible optical depth is low enough, the evaporative flux always exceeds the condensation flux.  This means that there is no snow line {\it per se} in the photosphere.  Second, although ice grains are nowhere stable, the evaporative flux may be so low that the lifetime of the grains will exceed the lifetime of the disk.  In such cases, the usual definition of snow line needs to be modified.  Finally, the radial distance from the star at which ice is stable will vary with height above midplane.  In such a case too, the concept of ``snow line" should be replaced by ``ice stability region".

\section{Effects of a Grain Size Distribution}
\begin{figure}
 \vspace*{2.0 cm}
\begin{center}
 \includegraphics[width=10cm]{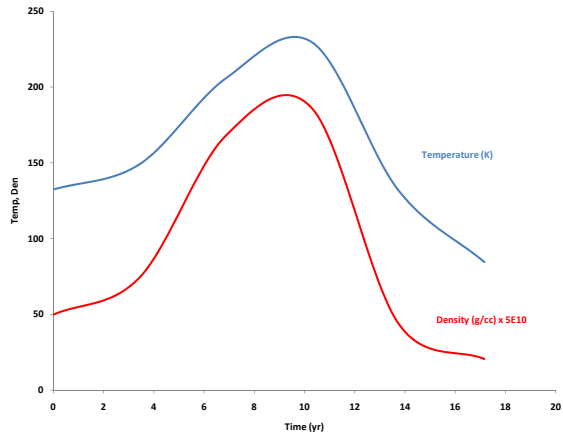}
 \vspace*{-3.0 cm}
 \caption{Temperature (blue) and density (red) profile in a density wave as a function of time.  Based on the work of Mayer (personal communication).}
   \label{dwave0}
\end{center}
\end{figure}

In addition to the effects of energy balance, there can also be effects due to exchange of energy (in the form of water vapor) between grains of different sizes.  When only one grain size is involved, a steady state with the background gas is reached when the rate of condensation from the background onto the grain is equal to the rate of evaporation off the grain.  If there is more than one grain size involved, it is possible, even in an optically thick region, for the grains to be at slightly different temperatures.  In such a case there will migration of ice from the hotter grain to the colder grain, and this migration will be moderated by the background vapor pressure as determined by the gas temperature.  This can lead to an interesting hysteresis phenomenon as demonstrated by the following example:

Gas disks can develop density waves (see, e.g. \cite[Meyer et al. 2002]{Meyeretal02}).  These waves cause a change in the gas pressure and temperature with time.  The changes are different in different parts of the wave, but a typical case is shown in Fig. \ref{dwave0} based on the work of Mayer (personal communication).   
\begin{figure}
 \vspace*{2.0 cm}
\begin{center}
 \includegraphics[width=10cm]{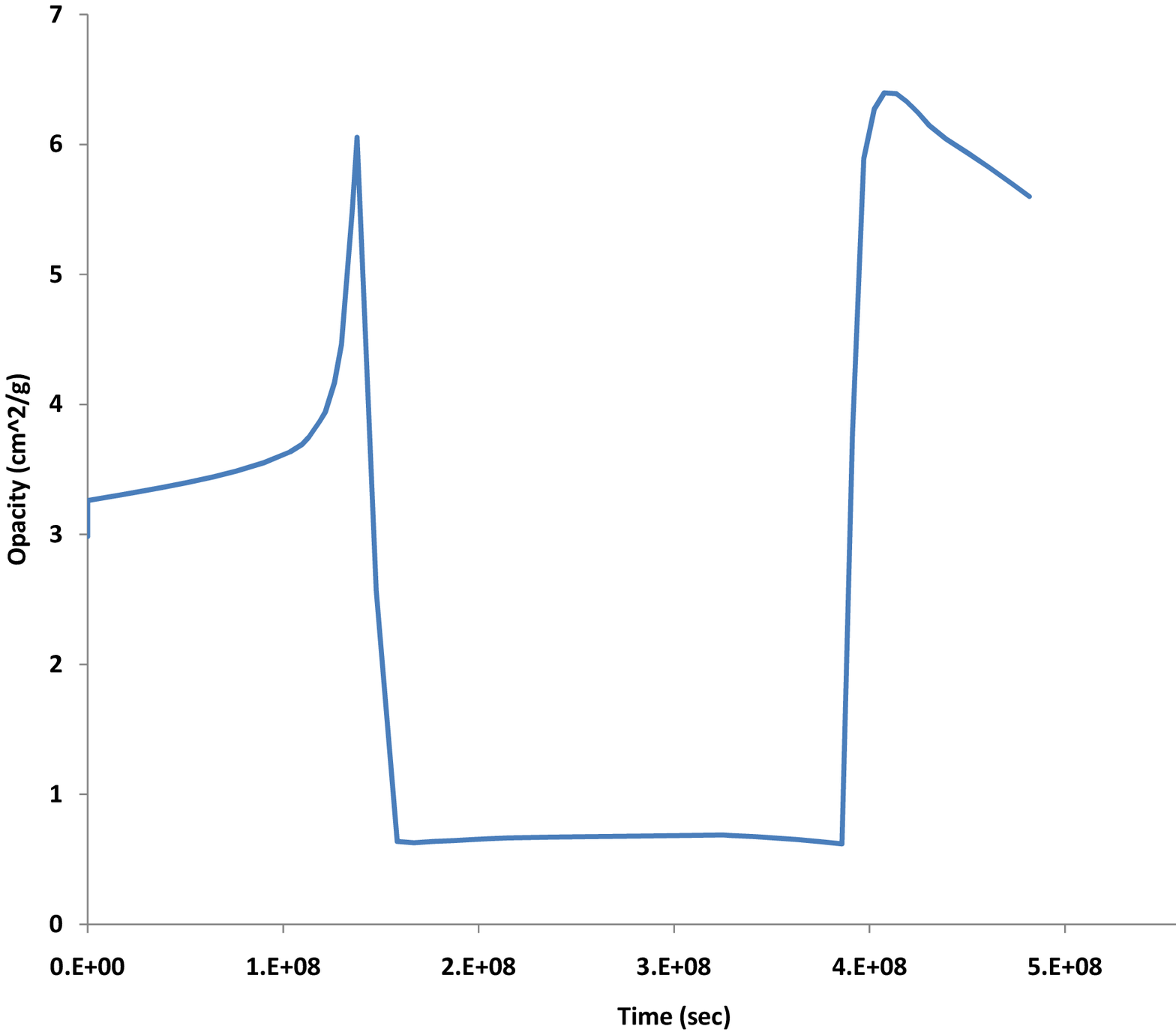}
\includegraphics[width=10cm]{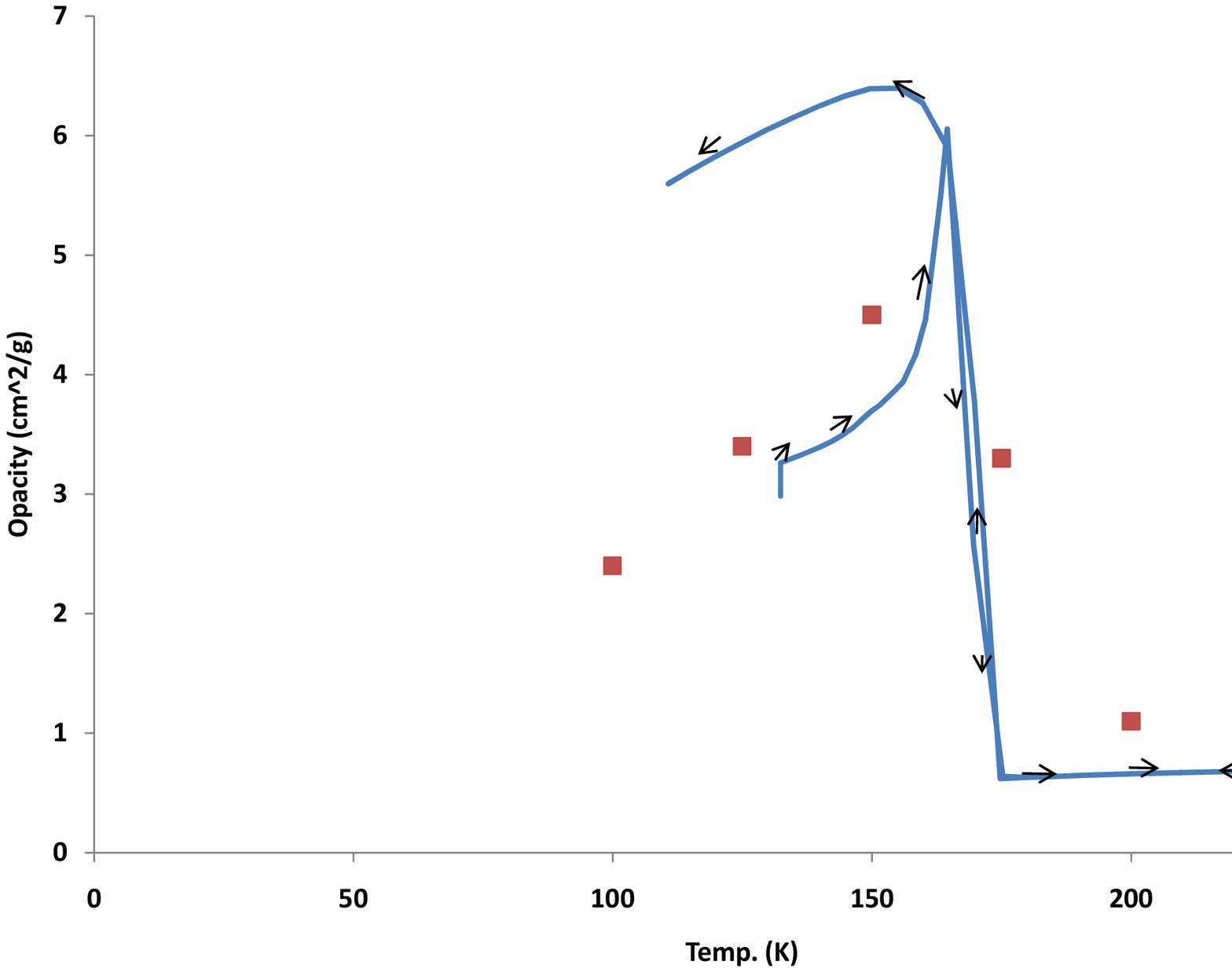} 
 \vspace*{-3.0 cm}
 \caption{The upper panel shows the opacity as a function of time for a grain size distribution as a density wave develops. The lower panel shows the opacity as a function of temperature.  The arrows indicate the progression as a function of time.  The squares are the grain opacities of \cite[Pollack et al. (1985)]{Pollacketal85}.}
   \label{dwave}
\end{center}
\end{figure}
Here the gas is assumed to have the solar ratio of water to silicon.  The silicon is in the form of grains with a power law size distribution.  In this particular case the number of grains was taken to vary with grain size, $a$, as $a^{-3.5}$, which is typical of such distributions.  The gas is originally cold and the water vapor is highly supersaturated.  As the density wave develops, the gas temperature rises from $\sim 135$\,K to $\sim 220$\,K and then returns to its original low value (Meyer, personal communication).

The upper panel shows the Rosseland mean grain opacity for this size distribution.  At $t=0$ only the silicate cores contribute, but immediately the supersaturated vapor condenses onto the cores and forms an ice shell, causing an almost instantaneous rise in the opacity.  Since the flux is mainly from the gas onto the grain, the flux is nearly independent of the size and temperature of the grain, and each grain grows an ice shell of nearly the same thickness.  However, since the grain temperature is determined by the interplay of several processes of heating and cooling, and some of these processes depend on the grain size, the different grains sizes will have slightly different temperatures.  Differences in grain composition can also lead to such temperature differences.  Since the vapor pressure is very sensitive to temperature, even slight differences in temperature can lead to significant differences in evaporation rates.  The details depend on the ambient conditions, but for the case shown in the figure, the larger grains in the distribution are slightly cooler than the smaller ones.  Because of this, the rate of evaporation from the smaller grains is slightly higher than the rate of evaporation from the larger grains.  Since the pressure of water vapor in the disk is fixed by the temperature of the disk gas, the rate of condensation onto the grains is independent of grain size.  The result is a migration of ice from the smaller grains to the larger ones, with the rate of this migration being moderated by the temperature of the background gas.  At low temperatures ($T\apprle 135$\,K) the migration is so slow that it can be safely ignored, but at higher temperatures it quickly becomes important.

As the density wave develops, the gas heats up and the migration of ice speeds up.  As can be seen from the upper planel in Fig.\,\ref{dwave}, the rate of migration increases with temperature (time) until the temperature becomes high enough so that the ice shells of even the coolest grains are lost.  At this point the opacity drops to that of the pure silicate cores.  When the gas temperature drops again, there is once more a rapid recondensation of water vapor onto the silicate cores, but because the gas temperature is still relatively high, the migration of ice to the larger grains is very quick.  As the gas continues to cool, the new ice distribution remains fixed.  The result is a sort of hysteresis effect, where although the temperature of the gas returns to its original value, the opacity changes.  This is shown more explicitly in the bottom panel of Fig.\,\ref{dwave}, where the opacity as a function of temperature is shown.  The arrows show the direction of time flow in this simulation, and the red squares show the opacities of \cite[Pollack et al. (1985)]{Polletal85}.  As can be seen, this hysteresis effect can increase the opacity at a given temperature by a factor of two in some cases.

\section{Conclusions}
The snow line is an important concept that influences our understanding both of the composition of solar system bodies and the processes that lead to their formation.  In addition, the position of the snow line has consequences for the opacity of the disk and its evolution.  However, the position of the snow line is not merely a function of the temperature of the disk gas.  It also depends on the density of water vapor in the disk, and on the details of the radiation field, as well as on the size and composition of the ice grains.  As a result, it is not proper to talk about a ``snow line" {\it per se}, but of an ``ice stability region" ({\it ISR}).  The borders of this region will depend not only on distance from the central star, but also on the height above midplane.  The size distribution of ice grains in this {\it ISR} may vary from place to place, and may even show a time dependence through the hysteresis effect described above.  Theoretical predictions of where the borders of this {\it ISR} are located, as well as interpretations of observations of ice grains will need to base themselves on detailed modeling of the physics of such grains.

\end{document}